# Properties of three seismic events in September 2017 in the northern Korean Peninsula from moment tensor inversion

**Libo Han, Zhongliang Wu, Changsheng Jiang\*, Jie Liu**

Institute of Geophysics, China Earthquake Administration, 100081 Beijing, China
China Earthquake Networks Center, 100036 Beijing, China
\* Corresponding author, e-mail: Jiangcs@cea-igp.ac.cn

## ABSTRACT

Moment tensor inversion is conducted to characterize the source properties of the September 3, M6.3, the September 3, M4.6, and the September 23, M3.4 seismic events occurred in 2017 in the nuclear test site of DPRK. To overcome the difficulties in the comparison, the inversion uses the same stations, the same structural model, the same algorithm, and nearly the same filters in the processing of waveforms. It is shown that the M6.3 event is with predominant explosion component, the M4.6 event is with predominant implosion component, while the M3.4 event is with a predominant double couple component (~74%) and a secondary explosion component (~25%). The three seismic events are with a similar centroid depth. The double couple component of the M3.4 event shows a normal fault striking northeastward.

## INTRODUCTION

In September 2017 the northern Korean Peninsula dramatically caused the attention of the world by a series of seismic events of M6.3 and M4.6 on September 3, and M3.4 on September 23, respectively. Table 1 lists the parameters of such seismic events as reported by China Earthquake Networks Center (CENC). DPRK authorities announced that the M6.3 event was a "successfully conducted hydrogen-bomb test". The M4.6 event, some 8 minutes followed, was regarded as a collapse. Different from previous similar seismic events which have been studied in various aspects, for example the yield and the accurate/precise location (*e.g.* Kim and Richards, 2007; Kvaerna *et al*., 2007; Bonner *et al*., 2008; Koper *et al*., 2008; Kim *et al*., 2009; Schlittenhardt *et al*., 2010; Shin *et al*., 2010; Wen and Long, 2010; Chun *et al*., 2011; Rougier *et al*., 2011; Pasyanos *et al*., 2012; Zhao *et al*., 2012; Zhang and Wen, 2013; Zhao *et al*., 2014), properties of, and relation between, these three seismic events, especially the source property of the last small event, raises scientific problems to be discussed by seismological data.

Shortly after the occurrence of the M3.4 event on September 23, some results were reported by the Weibo/WeChat (Chinese version of Twitter/Facebook). The result from the P/S spectrum ratio indicated that this event is not within the cluster of the UNEs (http://www.igg.cas.cn/xwzx/kyjz/201709/t20170923_4863959.html), but for the source property of this event there were two different opinions, namely natural earthquake, or collapse (http://wemedia.ifeng.com/30848756/wemedia.shtml; Chen Huizhong, 2017, WeChat communication, 2017/09/23; Liu Jie, 2017, *ibid*; Su Jinrong, 2017, *ibid*; Peng Zhigang, 2017, *ibid*; Zhao Lianfeng, 2017, *ibid*. Reports from different experts varied with time, yet from the related media coverage, "traces" of such discrepancy can be seen clearly). Analysts of China Earthquake Networks Center (CENC) insisted that the event is associated with an explosion (http://www.cenc.ac.cn/cenc/dzxx/336221/index.html). Due to the concern of the security situation of the Korean Peninsula as well as the potential environmental effects associated with the UNE detonation (Console and Nikolaev, 1995), the properties of this seismic event, as well as the discrepancies between different agencies, caused the attention, and even critical comments, in the public.

The spirit of forensic seismology (Bowers and Selby, 2009), like the works associated with jurisdiction, is to exploit the data available, even if with limited coverage and quality, to get some evidences of, at least clues to, the reliable and/or persuasive conclusions. Characterization of the three seismic events has obvious limit from the data available due to poor geographic coverage of seismic stations, distance from the recording stations to the epicenter, the small magnitude of the last seismic event, and the mixing of the seismic waveform of the second event with that of the first one. However, by some semi-quantitative, *e.g.* comparative, analysis, yet some qualitative conclusions could be obtained, which may help in the judgment of the properties of these seismic events. In this research letter we try to discuss this question in the perspective of seismic moment tensor, which seems not mentioned in the discussion among different agencies.

## FULL MOMENT TENSOR INVERSION

Characterization of seismic events (including underground nuclear tests) in the perspective of seismic moment tensor, as well as its theoretical discussion, has a long history (*e.g.* Wu and Chen, 1996; Dufumier and Rivera, 1997; Dreger and Woods, 2002; Ford et al., 2009a, 2009b). Moment tensor retrieval has its uncertainties, sometimes large due to the structural model used in the inversion, the seismic phases selected, and filter in use before the pre-processing (Šílený *et al*., 1996). Our inversion also suffers from the above-mentioned limitation. Considering this difficulty, we concentrate our attention on two aspects in the analysis: One is the predominant component in the resultant moment tensor, which is an indication of the source property, namely an explosion, an implosion, or a double couple; The other is the relative difference of the centroid depths of these three events, keeping in mind that there might be large systematic bias due to the structural model used.

In this study, the Green functions were calculated with the frequency-wavenumber (F-K) technique (Zhu and Rivera, 2002) with the ak135 1-D velocity model (Kennett *et al*. 1995). The "Cut and Paste (CAP)" algorithm for inverting centroid moment tensor, firstly developed by Zhao and Helmberger (1994) and then modified by Zhu and Helmberger (1996), is adopted for each seismic event. Advantage of this algorithm is its capability of efficiently reducing the uncertainties from unknown velocity structures by simply separating the entire seismograms into Pnl and surface wave segments and allowing for relative time shift between them. Currently, the method is widely applied to determine focal mechanisms of small-to-moderate earthquakes and has been used in the routine analyses of China Earthquake Administration (CEA) for earthquake emergency response. Previous results show that this algorithm works not only for natural earthquakes but also for other types of seismic events, with reliable characterization of the isotropic component of the seismic source (*e.g*. Ross *et al*., 2015). This development, the original CAP which works for double couple to generalized CAP (gCAP) which works for full moment tensor, was made by Zhu and Ben-Zion (2013) based on the formulation of Chapman and Leaney (2011). Decomposing a moment tensor into an isotropic component and a deviatoric component, Zhu and Ben-Zion (2013) introduced a dimensionless parameter ζ to quantify the relative strength of the isotropic component varying from -1 (implosion) to 1(explosion). This parameter is used in this study to characterize the properties of the three seismic events under discussion.

## DATA USED FOR ANALYSIS

Figure 1 shows the seismic stations used in the waveform inversion. To overcome the difficulties in the inversion, we purposely use the same stations, the same structure model, and nearly the same filters for pre-processing of the seismic waveforms.

Seismic recordings are from the data center of the China National Seismograph Network (Zheng *et al*., 2010). The seismic stations, located to the China-DPRK border and belonging to the regional networks of Jilin Province and Liaoning Province, respectively, use broadband seismographs, with flat frequency response for various seismographs 60 sec to 40 Hz (CMG-3ESPC and BBVS-60 seismograph), 120 sec to 40 Hz (CTS-1 seismograph), and 360 sec to 40 Hz (JCZ-1 seismograph), respectively. For the largest M6.3 event, a band-pass filter 50 sec to 0.12 Hz is adopted, while for the M4.6 event and the M3.4 event, they are both 20 sec to 0.12 Hz. This passband makes the source time function of the UNEs as well as small earthquakes, as can be resolved by the seismic recordings several tens of kilometers away, approximates a pulse with short duration. In this study, it is a triangle with duration 0.1 sec.

Due to the origin time of the M4.6 event which makes the waveform of this event mixed with the coda of the M6.3 event, and the magnitude of the M3.4 event which produced only weak records with low signal-to-noise ratio (SNR) at the stations available with epicentral distances from about 70 km to about 320 km, the P waves of these two events are not clear. As a result, only surface wave segments are used in the inversion. Because Pnl segments are not included, the CAP tactics seemingly plays a minor role. However, because three component seismograms are used in the inversion, in which the vertical/radial and tangential components can have different time shifts in the CAP practice, the CAP tactics still contributes significantly to overcome the difficulty caused by the uncertainties in the structural model.

## CHARACTERIZATION OF SOURCE PROPERTIES

To some extent, the exercises conducted in this study has dual missions, that to use the M6.3 event and the M4.6 event as the "calibration" events to test the reliability of the inversion, and to use the inversion to characterize the M3.4 event.

Figures 2~4 show the full moment tensor inversion results for these three seismic events, from which it can be seen that the synthetic seismograms and the observed waveforms have a good fit. Table 2 lists the inversion results, especially the relative weight of the isotropic (ISO) and double couple (DC) components. The dimensionless parameter $\zeta$ is used to quantify the source properties as per whether the seismic event

is an explosion (ζ near to 1) or a collapse (ζ near to −1). Seen from the moment tensor inversion result, the M6.3 event is characterized by a predominant explosion (ζ = 0.85), and the M4.6 event followed is featured by a predominant implosion (ζ =−0.92), which is qualitatively consistent with the widely accepted conclusions, indicating the reliability of the inversion algorithm. The M3.4 event, which is the focus of discussion, seems to have a predominant double couple component (~74%) plus a secondary isotropic positive component (~25%). The double couple component, with strike 231°, dip 68° and rake −122°, indicates a normal faulting. Even if considering the uncertainties of the moment tensor inversion (Šílený *et al.*, 1996), the isotropic component (ζ = 0.50) seems by no means negligible, and what is important is that it is not a collapse.

Comparing to the ISO or the DC component, the CLVD component associated with the three seismic events is small, being an order of magnitude smaller than the ISO and the DC component. Considering the uncertainties of the inversion, it is hard to get any conclusion about the portion and significance of the CLVD component.

Figure 5 shows the $m_b$:$M_S$ plot for the northern Korean Peninsula explosions with comparison with other UNEs and earthquakes, adding the recent 4 explosions since 2013 of the northern Korean Peninsula to Figure 2 of Bowers and Selby (2009), with both $M_S$ and $m_b$ results from USGS. According to previously conducted theoretical analysis, the effect of tensile failure on $M_S$ is to enhance the explosion-like characteristics on a plot of $m_b$:$M_S$ (Patton and Taylor, 2008). This result suggests that the success of the traditional $m_b$:$M_S$ discriminant results from the fact that nuclear tests were conducted under containment practices for which tensile failure is ubiquitous, while the DPRK nuclear test of 2006 is a harbinger of poor $m_b$:$M_S$ performance when tensile failure is completely suppressed. From Figure 5 it can be seen that for the M6.3 event, the $m_b$:$M_S$ discriminant works well. One of the plausible conjectures might be that the good $m_b$:$M_S$ performance could be explained by the existence of the tensile spall associated with the detonation, and the M4.6 event (being an implosion) could be considered as the collapse of the spall itself.

A more complicated problem is associated with the hypocentral/centroid depth. Figure 6 shows the waveform fit error as a function of trial centroid depth of the three seismic events. Note that due to the structural model used in the CMT inversion, the results for the three events may have systematic bias. To overcome this difficulty for the comparison, the same stations, same seismic phases, same structural model, and approximately the same pass-bands are used in the inversion, From the figure it can be observed that the centroid depths of the three events are nearly the same, albeit the ranges of uncertainty for the M4.6 and the M3.4 event are apparently larger, being about ±0.5 km and ±1 km, respectively. Yet "being nearly the same" is the furthest we may reach, keeping in mind that the explicit value ~2.5 km is not reliable. Considering the result of previous studies (Rougier *et al*., 2011), probably we can say that 2.5 km only represents the order of magnitudes of the true depth which is estimated as some hundred meters. Accurate and precise location of these three seismic events is beyond the scope of this study. Taking the previously published results (either via scientific journals or via new media) we simply assume that the location of these three events are "seismologically" the same, that is, within a range whose size is comparable to the predominant wavelength of the waveforms in use. On the other hand, the similarity of the centroid depths of the three events may still provide some hints to their relative locations and the possible causes of them as well.

**DISCUSSION AND CONCLUSIONS**

In this study we use moment tensor inversion to characterize the source properties of the September 3, M6.3, the September 3, M4.6, and the September 23, M3.4 seismic events occurred in 2017 in the nuclear test site of DPRK. We use the same stations, the same structural model, the same algorithm, and nearly the same filters in the pre-processing of waveforms to facilitate the comparison. It is shown that the M6.3 event is with predominant explosion component, and the M4.6 event is with predominant implosion component, which is consistent with the well accepted conclusions up to now. The M3.4 event, which is the focus of interest in this study stimulated to much extent by the debate on its source property, exhibits a predominant double couple component (~74%) and a secondary explosion component (~25%). Currently, we do not have sound seismological data which is sufficient to constrain the origin of this explosion component, but this isotropic expansion component might be an explanation why the analysts of China Earthquake Networks Center (CENC), in their early stage of quick epicenter report, noted that the event was "suspicious to be an explosion".

Generally speaking, the currently used seismic moment tensor inversion algorithms can reveal the predominant component of a moment tensor, and thus can help in judging whether it is mainly an explosion, a collapse, or a double couple (*e.g.* Ford *et al*., 2008; Ortega *et al*., 2014), while the secondary component may be problematic due to the resolution of the inversion (*e.g.* Šílený *et al*., 1996). This may also cause some problems when the predominant and the secondary components are comparative, that the source parameters of the double couple (that is, the strike, dip, and rake angle, as well as its scalar seismic moment) might be problematic. The double couple component of the M3.4 event shows a normal faulting mechanism striking northeastward. However, considering the above mentioned complexities, caution has to be taken that it would be on a shaky ground if the comparison between this focal mechanism and local stress field were conducted. Similar caution is valid for the conclusion that the three seismic events are with a similar centroid depth.

## ACKNOWLEDGMENTS

Thanks to Prof. Zheng Guoguang of China Earthquake Administration (CEA) for stimulating discussion, and to Prof. Qiao Sen of the Institute of Geophysics, CEA, and Dr. Yu Shuming of the Department for Earthquake Monitoring and Prediction, CEA, for comments and suggestion in improving the manuscript. The CAP algorithm was provided by, and guided for application by Prof. Lupei Zhu of St. Louis University. Discussion with Dr. Zhang Ling of Liaoning Province Earthquake Agency helped much to the writing of the paper.

underground nuclear explosions, *Pure appl. Geophys*. **147**, 357-366.

Zhang, M., and L. Wen (2013). High-precision location and yield of North Korea's 2013 nuclear test, *Geophys. Res. Letts.* **40**, 2941-2946.

Zhao, L. F., X. B. Xie, W. M. Wang, and Z. X. Yao (2012). Yield estimation of the 25 May 2009 North Korean nuclear explosion, *Bull. Seismol. Soc. Am.* **102**, 467-478.

Zhao, L. F., X. B. Xie, W. M. Wang, and Z. X. Yao (2014). The 12 February 2013 North Korean underground nuclear test, *Seismol. Res. Lett*. **85**, 130-134.

Zhao, L. S., and D. V. Helmberger (1994). Source estimation from broadband regional seismograms, *Bull. Seismol. Soc. Am.* **84**, 91-104.

Zheng, X. F., Z. X. Yao, J. H. Liang, and J. Zheng (2010). The role played and opportunities provided by IGP DMC of China National Seismic Network in Wenchuan earthquake disaster relief and researches, *Bull. Seismol. Soc. Am.* **100**, 2866-2872.

Zhu, L., and D. V. Helmberger (1996). Advancement in source estimation techniques using broadband regional seismograms, *Bull. Seismol. Soc. Am.* **86**, 1634-1641.

Zhu, L., and L. A. Rivera (2002). A note on dynamic and static displacements from a point source in multilayered media, *Geophys. J. Int.* **148**, 619-627.

Zhu, L., and Y. Ben-Zion (2013). Parametrization of general seismic potency and moment tensors for source inversion of seismic waveform data, *Geophys. J. Int*. **194**, 839-843.

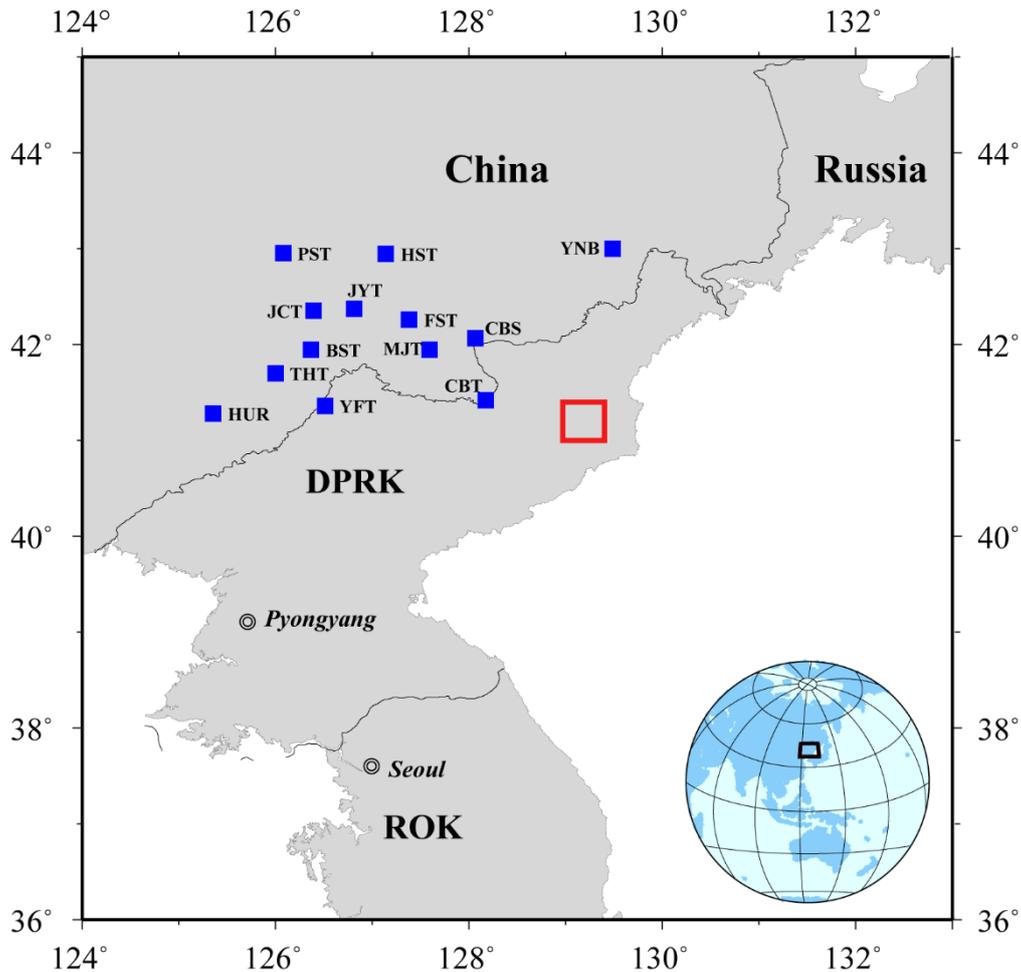

**Figure 1**. Map of the DPRK nuclear test site (red open square). The regional seismic stations used in the waveform inversion are marked as blue solid squares. Inset indicates this map in a global view. Stations shown in the figure are: Baishan (BST, Jilin Province, BBVS-60 seismograph), Changbai (CBT, Jilin Province, CMG-3ESPC seismograph), Changbaishan (CBS, Jilin Province, CTS-1 seismograph), Fusong (FST, Jilin Province, CMG-3ESPC seismograph), Huanren (HUR, Liaoning Province, BBVS-60 seismograph), Hongshi (HST, Jilin Province, BBVS-60 seismograph) , Jinchuan (JCT, Jilin Province, BBVS-60 seismograph) , JinYu (JYT, Jilin Province, BBVS-60 seismograph), Manjiang (MJT, Jilin Province, BBVS-60 seismograph), Panshi (PST, Jilin Province, BBVS-60 seismograph), Tonghua (THT, Jilin Province, CTS-1 seismograph), Yanbian (YNB, Jilin Province, JCZ-1 seismograph), and Yunfeng (YFT, Jilin Province, BBVS-60 seismograph) with all the seismographs broadband. In figures 2~4 the station names are with suffix indicating the regional network the station belonging to, for example, Huanren station of the Liaoning Province Seismograph Network is marked as HUR.LN.

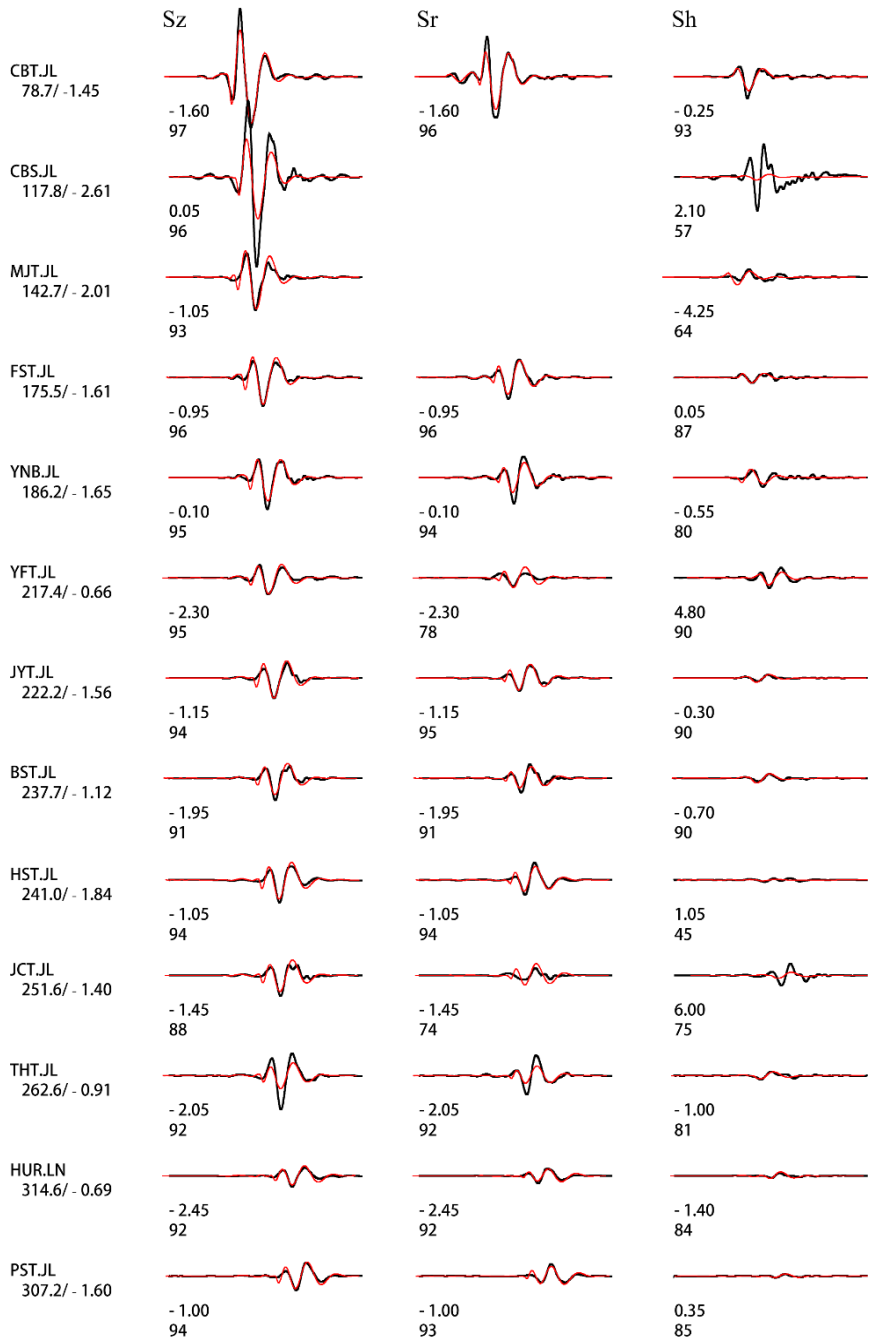

**Figure 2**. Waveform fitting in the moment tensor inversion for the Sep 3 M6.3 seismic event, with comparison between observed (black) and synthetic (red) seismograms. The three columns of seismograms are vertical, radial and tangential components of surface waves, respectively. The numbers below the station names are the epicenter distances in kilometers and overall time shifts in the CAP inversion. The first number below each seismogram are time shift between observed data and synthetics for best fitting (positive indicates synthetic is earlier than the observed), and the second number is the cross-correlation coefficient between synthetic and observed seismograms (in percentage).

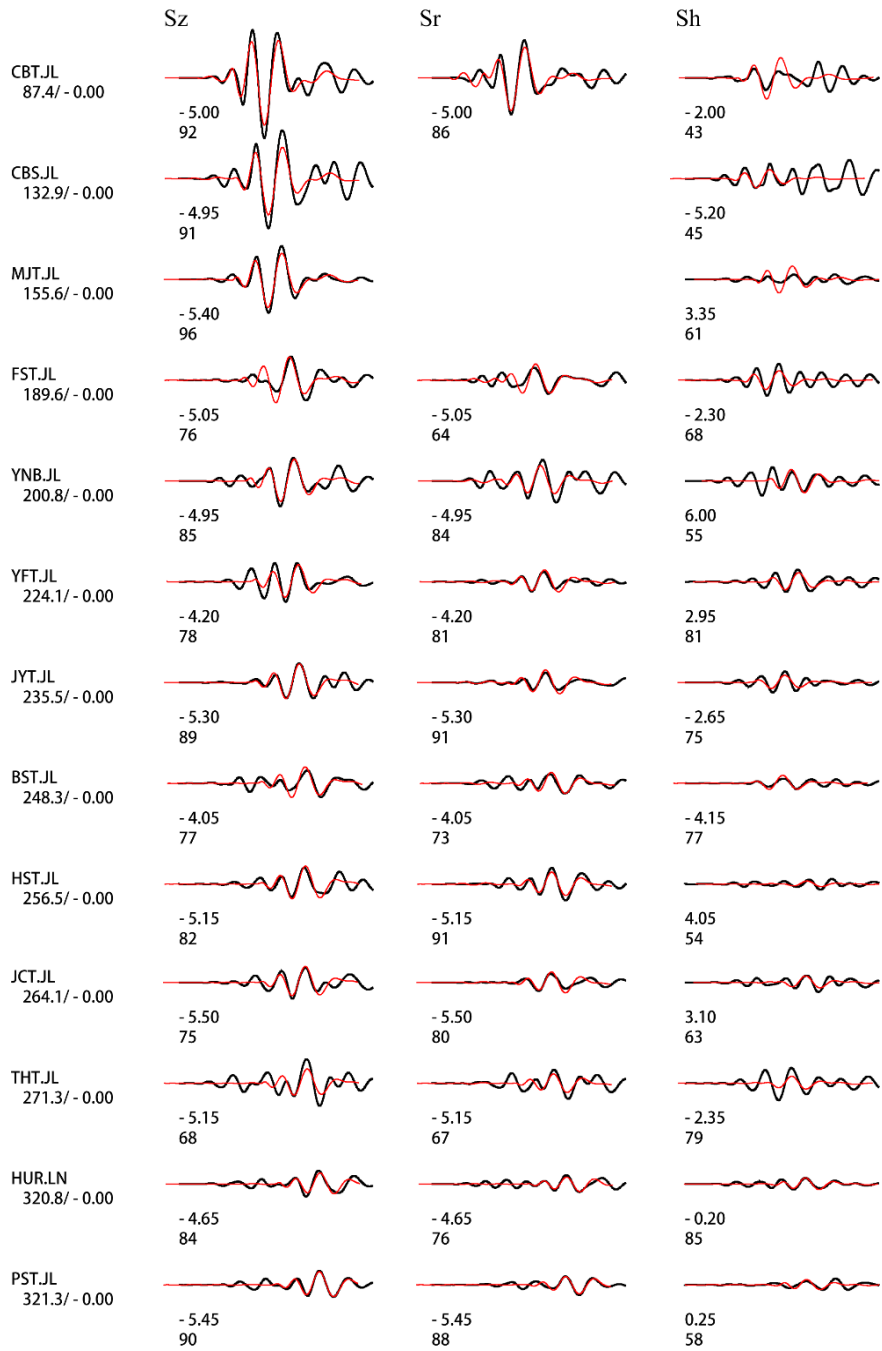

**Figure 3**. Same as Figure 2, for the Sep 3 M4.6 event.

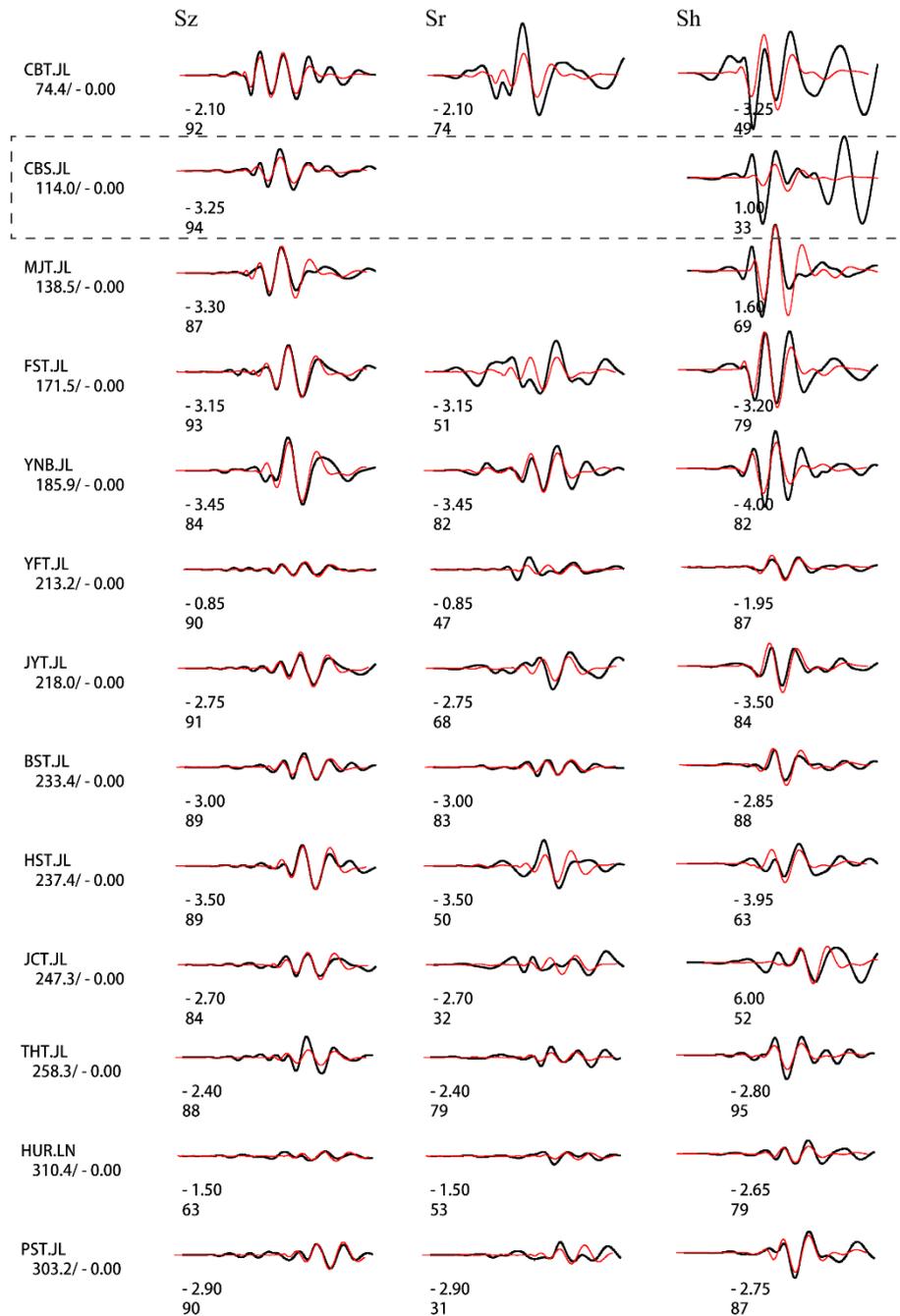

**Figure 4**. Same as Figure 2, for the Sep 23 M3.4 event. In the figure due to the abnormal large amplitude of the tangential component surface wave at the CBS.JL station, the synthetic and observed waveforms in the box use a different scale, with 1/3 of the other stations.

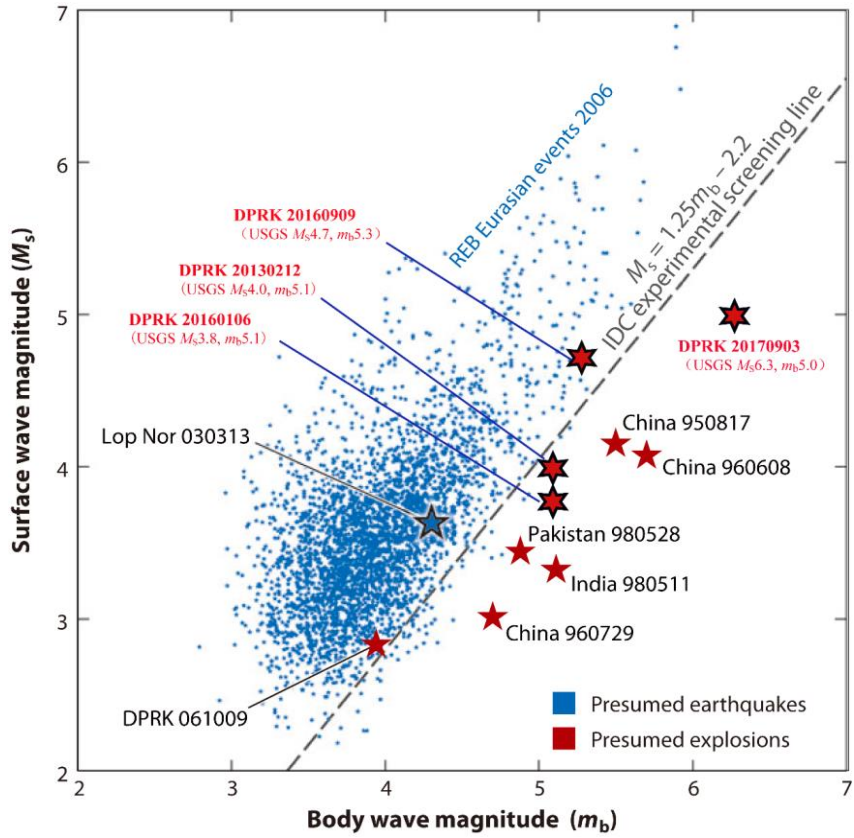

**Figure 5.** The $m_b$:$M_S$ plot for the northern Korean Peninsula explosions with comparison with other UNEs and earthquakes. The base map are modified from Figure 2 of Bowers and Selby (2009), adding the recent 4 explosions in the northern Korean Peninsula (marked by red text) with both $M_S$ and $m_b$ results from USGS.

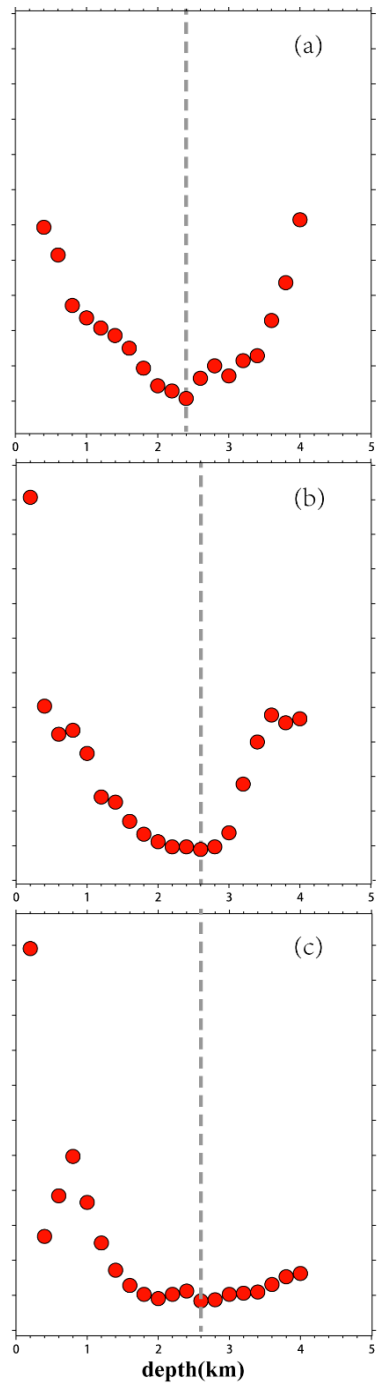

**Figure 6.** The waveform fit error (vertical axis, scaled) as a function of trial centroid depth (horizontal axis) of the three seismic events. (a) The Sep 3 M6.3 event; (b) The Sep 3 M4.6 event; (c) The Sep 23 M3.4 event. The vertical dashed lines indicate the "best" centroid depth. The error-depth curve also indicates the ranges of the "best" centroid depth.

# TABLE 1

# Seismic events under discussion in the present study

| Event | Origin time | | CENC Location | | Magnitude | |
|---|---|---|---|---|---|---|
| | Date (YYYY/MM/DD) | UTC Time (HH:MM:SS) | Latitude (°N) | Longitude (°E) | CENC | USGS |
| 1 | 2017/09/03 | 03:30:01 | 41.35 | 129.11 | $M_S$6.3 | $m_b$6.3 |
| 2 | 2017/09/03 | 03:38:31 | 41.21 | 129.18 | $M_S$4.6 | $M_L$4.1 |
| 3 | 2017/09/23 | 08:29:16 | 41.36 | 129.06 | $M_L$3.4 | $M_L$3.5 |

# TABLE 2

## Main moment tensor elements of the seismic events

| Event | Date (YYYY/MM/DD) | UTC Time (HH:MM:SS) | ζ | ISO (%) | DC (%) | CLVD (%) | Centroid depth (km) |
|---|---|---|---|---|---|---|---|
| 1 | 2017/09/03 | 03:30:01 | 0.85 | 72.3 | 26.6 | 1.1 | 2.4 |
| 2 | 2017/09/03 | 03:38:31 | -0.92 | -84.6 | 14.8 | 0.6 | 2.6 |
| 3 | 2017/09/23 | 08:29:16 | 0.50 | 25.0 | 73.9 | 1.1 | 2.6 |

Note: In the table the ζ-value (Zhu and Ben-Zion, 2013), a dimensionless parameter quantifying the relative strength of the isotropic component, varying from -1 (implosion) to 1 (explosion), is provided.